# Capacity Bounds for Wireless Ergodic Fading Broadcast Channels with Partial CSIT

Reza K. Farsani[1]

Email: reza_khosravi@alum.sharif.ir

*Abstract:* **The two-user wireless ergodic fading Broadcast Channel (BC) with partial Channel State Information at the Transmitter (CSIT) is considered. The CSIT is given by an arbitrary deterministic function of the channel state. This characteristic yields a full control over how much state information is available, from perfect to no information. In literature, capacity derivations for wireless ergodic fading channels, specifically for fading BCs, mostly rely on the analysis of channels comprising of parallel sub-channels. This technique is usually suitable for the cases where perfect state information is available at the transmitters. In this paper, new arguments are proposed to directly derive (without resorting to the analysis of parallel channels) capacity bounds for two-user fading BC with both common and private messages based on the existing bounds for the discrete channel. First, a capacity inner bound is proposed for the channel by choosing an appropriate signaling scheme for the Marton's achievable rate region. Then, a novel approach is developed to adapt and evaluate the well-known UV-outer bound for the Gaussian fading channel using the entropy power inequality. Our approach indeed sheds light on the role of broadcast auxiliaries in the fading channel. It is shown that the derived inner and outer bounds coincide for the channel with perfect CSIT as well as for some special cases with partial CSIT. Our bounds are also directly applicable to the case without CSIT which has been recently considered in several papers. Next, the approach is developed to analyze for the fading BC with secrecy. In the case of perfect CSIT, a full characterization of the secrecy capacity region is derived for the channel with common and confidential messages. This result completes a gap in a previous work by Ekrem and Ulukus. For the channel without common message, the secrecy capacity region is also derived when the transmitter has access only to the degradedness ordering of the channel.**

## I. Introduction

In wireless communication networks, due to the mobility of users, the channel from the transmitters to each receiver is corrupted by time-varying (multiplicative) fading coefficients in addition to the additive noise. The fading networks have been widely studied in communication theory, however, there exist still many open problems regarding fundamental limits of communications in these scenarios. One of the basic networks which is of great importance both from practical and theoretical viewpoints is the Broadcast Channel (BC). For the wireless ergodic fading BC the capacity region is only known when perfect channel state information is available at both the transmitter and the receivers [1]. The assumption of perfect Channel State Information at the Transmitter (CSIT) is rather restrictive because in many practical applications it is not possible to provide this information for the transmitter in a timely manner. In the case of without CSIT, the two-user ergodic fading BC has been considered in several papers. In [2] an achievable rate region was proposed for the channel based on superposition coding. The paper [3] considers the one-sided channel where one of the users has a constant (non-fading) channel. In [4], a layered erasure channel was proposed to approximate the Gaussian fading channel and then by taking insights from the erasure model, inner and outer bounds were derived for the fading BC without CSIT which are within to a constant gap of each other for all fading distributions. Also, an improved constant gap result was obtained for the channel in [5]. In the paper [6], the channel with discrete-valued (belonging to a finite subset of real numbers) fading coefficients was considered and outer bounds were derived using Costa's Entropy Power Inequality (EPI). However, despite its practical importance, the capacity region of the fading BC without CSIT is still an open problem.

In this paper, we consider the two-user fading BC with partial CSIT. The CSIT is given by an arbitrary deterministic function (potentially discrete-valued) of channel state. The main benefit of such a model for CSIT, which was considered previously in [7] and [8] for the fading multiple access channels, is that it provides a full control over how much state information is available from perfect to no information. In literature capacity derivations for wireless ergodic channels, specifically fading BCs, mostly rely on the analysis of channels comprising of parallel sub-channels [1, 9-15]. By this approach the ergodic capacity region was established in [1, 9-13] for different multi-user fading channels with perfect CSIT. In fact, this technique that fading channels are treated based on parallel channels is usually suitable for the cases where perfect state information is available at the transmitters. Moreover, it is no more applicable for analyzing scenarios such as fading interference channels that are not separable into parallel sub-channels [16-17]. In this paper, we present novel arguments to directly derive (without resorting to the analysis of parallel channels) capacity bounds for the two-user fading BC with partial CSIT with both common and private messages based on the existing bounds for the discrete channel. First, we propose a capacity inner bound for the channel by choosing an appropriate signaling scheme for the Marton's achievable rate region. We then establish an outer bound on the capacity region. We remark that one of the main challenges in analyzing the fading BC is to establish a capacity outer bound with satisfactory performance. Specifically, it has been a main focus in all the papers [2-6]. The reason is that capacity outer bounds for the BC typically include some auxiliary random variables and for the Gaussian fading channel (unlike the Gaussian channel with fixed channel gains) a naive application of the EPI to optimize over these auxiliaries fails. In [6], the authors indicate that conventional EPI is not directly applicable for analyzing the fading BC without CSIT and instead make use of Costa's EPI for this purpose. Also, the outer bound given in [4] for this channel is derived using a channel enhancement technique (which creates a degraded channel) and then the relations between mutual information and minimum mean square error [18] are used to optimize over its auxiliary random variable (whose role is less clear in the Gaussian fading channel [19, Conclusion]). Nonetheless, in this paper, we develop a novel and rather simple approach to adapt and evaluate the well-known UV-outer bound [20] for the Gaussian fading BC using the EPI. Our approach indeed sheds light on the role of broadcast auxiliaries in the fading channel. We next prove that our inner and outer bounds coincide for the channel with perfect CSIT. For the special case of the fading BC without common message, the result of [1] is thus recovered with a new and concise proof. The capacity region is also derived for some new special cases with partial CSIT. Our bounds are directly applicable to the case without CSIT, as well.

---
[1]Reza K. Farsani was with the department of electrical engineering, Sharif University of Technology. He is by now with the school of cognitive sciences, Institute for Research in Fundamental Sciences (IPM), Tehran, Iran.

Also, we develop our approach to analyze for the wireless ergodic fading BC with secrecy. In this scenario, a transmitter sends a common message and also two private messages to two receivers and wishes to keep each private message as secret as possible from the non-legitimate receiver. Special cases of this system have been previously considered in [9-12]. The derivations of all these papers rely on the analysis of fading channels using parallel channels. Also, all of them consider the fading channel with perfect CSIT. In this paper, we establish inner and outer bounds on the secrecy capacity region of the ergodic fading BC with partial CSIT for the general case where a common message and two confidential messages are transmitted. A key step in our analysis is to derive the outer bound. For this purpose, the outer bound established in [21] for the capacity-equivocation region of the discrete BC is exploited. We adapt this outer bound for the secrecy capacity region of the fading channel first and then optimize it over its auxiliary random variables using novel techniques. In the case of perfect CSIT, our inner and outer bounds coincide with each other, thus establishing a full characterization of the secrecy capacity region for the channel with both common and confidential messages. This result include all the ones derived in [9-12] as special cases. A gap in a previous work by Ekrem and Ulukus [11] is also completed. Clearly, in [11] Ekrem and Ulukus could find the secrecy capacity region of the parallel degraded BCs [11, Corollary 1] with both common and confidential messages, however, for the Gaussian fading channel the secrecy capacity region is given only for the channel without common message (in other words, for the Gaussian fading BC with both common and confidential messages the secrecy capacity region remains unresolved in [11]). For the channel without common message, we also establish the secrecy capacity region when the transmitter has access only to the degradedness ordering of the channel which is a more realistic assumption than perfect CSIT.

It should be noted that in this paper we consider the two-user BC; however, our approach is also applicable for other multi-user fading networks, specially the fading interference channels [23]. In general, the benefits of our approach in the analysis of fading channels can be summarized as follows:

- ✓ The approach is applicable to analyze stationary ergodic Gaussian fading channels with arbitrary fading statistics.
- ✓ The analysis is concise. The bounds for the fading channels are directly built upon the existing results for the corresponding discrete channel. Also, the outer bounds are optimized over auxiliary random variables by using a subtle application of the conventional EPI.
- ✓ The quality of CSIT becomes immaterial. We can analyze for how much state information is available at the transmitters from perfect to no information.
- ✓ The approach is applicable to analyze various fading network topologies regardless of that a given network is separable into parallel sub-channels or not. Specifically, in [23] by following the same approach, we derive the capacity region of the ergodic fading interference channel with partial CSIT to within one bit.

We also remark that this paper presents our approach for the derivation of capacity bounds. The explicit computation of the derived bounds is addressed in [3 2]. In the following, the channel model is defined in Section II and the main results are given in Section III.

Reza K. Farsani, 2013

## II. CHANNEL MODEL

In this paper, the following notations are used: The set of complex numbers, real numbers, and nonnegative real numbers are represented by $\mathbb{C}, \mathbb{R}$, and $\mathbb{R}_+$, respectively. The notation $\mathbb{E}[.]$ denotes the expectation operator. Given a statement $F$, the indicator function $\mathbb{1}(F)$ is equal to one if $F$ is true and zero otherwise. Also, for any real number $x$, the function $[x]_+$ is equal to $x$ if $x$ is nonnegative and zero otherwise. Finally, the deterministic function $\psi(x)$ is given by: $\psi(x) \equiv \log(1 + x), \ x \in \mathbb{R}_+$.

*Two-User Gaussian Fading Broadcast Channel (GFBC):*

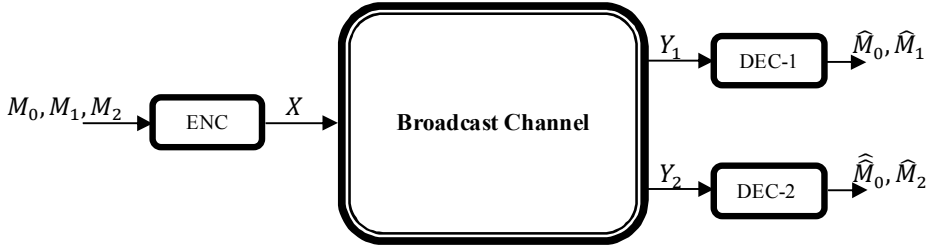

Figure 1. The two-user BC with common and private messages: $M_0$ is the common message and $M_1$ and $M_2$ are the private messages.

The two-user BC is a communication system where a transmitter broadcasts private messages and also a common message to two receivers (see Fig. 1). The Gaussian fading channel is given as follows:

$$\begin{cases} Y_{1,t} = S_{1,t}X_t + Z_{1,t} \\ Y_{2,t} = S_{2,t}X_t + Z_{2,t} \end{cases}, \quad t \geq 1$$

(1)

The sequence $\{X_t\}_{t \geq 1}$ denotes complex-valued transmitted signals by the transmitter, and $\{Y_{1,t}\}_{t \geq 1}$ and $\{Y_{2,t}\}_{t \geq 1}$ denote the received signals at the first and the second receivers, respectively. The sequences $\{Z_{1,t}\}_{t \geq 1}$ and $\{Z_{2,t}\}_{t \geq 1}$ represent additive noises each of which is an i.i.d. complex Gaussian random process with zero mean and unit variance. The channel state is denoted by $\{\boldsymbol{S}_t = (S_{1,t}, S_{2,t})\}_{t \geq 1}$ where the components $S_{1,t}$ and $S_{2,t}$ are (potentially correlated) complex-valued fading coefficients at the time instant $t$. In general, we suppose that the channel state is a stationary and ergodic random process with limited energy which varies in time according to an arbitrary (known) Probability Distribution Function (PDF). Nevertheless, some of the derived results hold only for channels with i.i.d. state process, which will be discriminated. It is also remarked that the state process of the channel is independent of the additive noises.

We assume that the state information is perfectly available at both receivers while the transmitter has access to it partially. The partial side information at the transmitter is prescribed by a deterministic function (potentially discrete-valued) of the channel state. Precisely, let $\xi(.)$ be a (arbitrary) deterministic function as:

$$\xi(.): \mathbb{C}^2 \to \mathcal{E}$$

where $\mathcal{E}$ is an arbitrary (potentially finite) set. At each time instant $t, t \geq 1$, the transmitter has access to $E_t = \xi(\boldsymbol{S}_t)$ where $\boldsymbol{S}_t$ is the current state of the channel. Below the encoding and decoding schemes are described in details.

*Encoding and decoding schemes:* For the two-user Gaussian fading BC (1), given a natural number $n$ and a triple $(R_0, R_1, R_2) \in \mathbb{R}_+^3$, a length-$n$ code $\mathfrak{C}^n(R_0, R_1, R_2)$ with a common message $M_0$ and two private messages $M_1$ and $M_2$ uniformly distributed over the sets $\{1, \ldots, 2^{nR_0}\}, \{1, \ldots, 2^{nR_1}\}$, and $\{1, \ldots, 2^{nR_2}\}$, respectively, consists of the following:

- ✓ A set of encoder mappings $\{\mathfrak{E}_t\}_{t=1}^n$ with:

$$\mathfrak{E}_t: \{1, \ldots, 2^{nR_0}\} \times \{1, \ldots, 2^{nR_1}\} \times \{1, \ldots, 2^{nR_2}\} \times \mathcal{E} \to \mathbb{C}$$

where the produced complex-valued signals are as $X_t = \mathfrak{E}_t(M_0, M_1, M_2, E_t), t = 1, \ldots, n$. The transmitter is subject to an average power constraint $P \in \mathbb{R}_+$, i.e.,

$$\frac{1}{n}\mathbb{E}[\sum_{t=1}^{n}|X_t|^2] \leq P$$

✓ Two decoder mappings $\mathfrak{D}_1(.)$ and $\mathfrak{D}_2(.)$ with:

$$\mathfrak{D}_i: \mathbb{C}^n \times \mathbb{C}^{2n} \to \{1, \ldots, 2^{nR_0}\} \times \{1, \ldots, 2^{nR_i}\}, \quad i = 1, 2$$

where the detected messages are as $(\widehat{M}_0, \widehat{M}_i) = \mathfrak{D}_i(Y_i^n, \boldsymbol{S}^n)$.

The triple $(R_0, R_1, R_2)$ represents the rate of the code. The average error probability of decoding is also given by:

$$P_e^{\mathfrak{C}^n} \triangleq P\left(\bigcup_{i=1,2}\{\mathfrak{D}_i(Y_i^n, \boldsymbol{S}^n) \neq M_i\}\right)$$

**Definition:** *For the two-user Gaussian ergodic fading BC (1), a triple $(R_0, R_1, R_2) \in \mathbb{R}_+^3$ is said to be achievable if there exist a sequence of codes $\mathfrak{C}^n(R_0, R_1, R_2)$ with $P_e^{\mathfrak{C}^n} \to 0$ as $n \to \infty$. The capacity region is the closure of the set of all achievable rates.*

For the channel with confidential messages, the secrecy level of the code is measured by normalized equivocation rates, i.e., $\frac{1}{n}H(M_1|Y_2^n, \boldsymbol{S}^n)$ and $\frac{1}{n}H(M_2|Y_1^n, \boldsymbol{S}^n)$. Accordingly, the secrecy capacity region is given as follows.

**Definition:** *For the two-user Gaussian ergodic fading BC (1) with secrecy, a triple $(R_0, R_1, R_2) \in \mathbb{R}_+^3$ is said to achievable if there exist a sequence of codes $\mathfrak{C}^n(R_0, R_1, R_2)$ with:*

$$\begin{cases} \lim_{n \to \infty} P_e^{\mathfrak{C}^n} = 0 \\ R_1 \leq \liminf_{n \to \infty} \frac{1}{n} H(M_1|Y_2^n, \boldsymbol{S}^n) \\ R_2 \leq \liminf_{n \to \infty} \frac{1}{n} H(M_2|Y_1^n, \boldsymbol{S}^n) \end{cases}$$

*The secrecy capacity region is the closure of the set of all achievable rates.*

Now, we are at the point to present our main results.

## III. MAIN RESULTS

In this section, we first propose a capacity inner bound for the two-user fading BC (1) with common message. This bound is derived by choosing an appropriate signaling scheme for the Marton's achievable rate region [24]. Next, we show that how one can adapt the structure of the so-called UV-outer bound [20] to be applicable for the Gaussian fading channel with stationary state process. Then, we present novel arguments to explicitly evaluate the derived outer bound. We also explore special cases where the inner and the outer bounds coincide which yield the capacity region. Finally, we follow our approach to derive bounds on the secrecy capacity region of the channel.

We begin by presenting the Marton's achievable rate region [24] adapted for the fading channel. Note that in what follows, the random variable $\boldsymbol{S} = (S_1, S_2) \in \mathbb{C}^2$ with a given PDF $P_{\boldsymbol{S}}(\boldsymbol{s})$ denotes the channel state, and $E = \xi(\boldsymbol{S})$ denotes the partial side information at the transmitter.

**Lemma 1)** *The rate region $\mathfrak{R}_i^{Marton \to GFBC}$ defined below constitutes an inner bound on the capacity region of the two-user Gaussian fading BC (1) with common message.*

$$\mathfrak{R}_i^{Marton \to GFBC} \triangleq \bigcup_{\substack{P_{WUV}, \\ f(.): X = f(W,U,V,E) \\ \mathbb{E}[|X|^2] \leq P}} \begin{Bmatrix} (R_0, R_1, R_2) \in \mathbb{R}_+^3 : \\ R_0 + R_1 \leq I(W, U; Y_1 | \mathbf{S}) \\ R_0 + R_2 \leq I(W, V; Y_2 | \mathbf{S}) \\ R_0 + R_1 + R_2 \leq I(U; Y_1 | W, \mathbf{S}) + I(W, V; Y_2 | \mathbf{S}) \\ \qquad \qquad -I(U; V | W) \\ R_0 + R_1 + R_2 \leq I(W, U; Y_1 | \mathbf{S}) + I(V; Y_2 | W, \mathbf{S}) \\ \qquad \qquad -I(U; V | W) \\ 2R_0 + R_1 + R_2 \leq I(W, U; Y_1 | \mathbf{S}) + I(W, V; Y_2 | \mathbf{S}) \\ \qquad \qquad -I(U; V | W) \end{Bmatrix}$$

(2)

*Proof of Lemma 1)* As mentioned, this achievable rate region is derived based on the Marton's coding strategy [24]. The messages are encoded by codewords generated (independent of the side information) based on the PDF $P_{WUV}(w, u, v)$ exactly similar to the Marton's scheme (see [25, Ch. 8]). The signaling at the transmitter for each time instant is given by a deterministic function $f(.)$ as $X = f(W, U, V, E)$, where $E$ is the side information available at the transmitter. The decoding procedure is also similar to the Marton's scheme, however, here the output signals for the first and the second users are considered as $(Y_1, \mathbf{S})$ and $(Y_2, \mathbf{S})$, respectively. ∎

We remark that optimizing the achievable rate region $\mathfrak{R}_i^{Marton \to GFBC}$ in (2) for the general case is difficult. Nevertheless, in the following we propose a useful signaling scheme for the channel based on this achievable rate region. As we see later, for the case where the transmitter knows the state perfectly, i.e., $E \equiv \mathbf{S}$, this scheme is indeed optimal.

**Proposition 1)** *Define the rate region $\mathfrak{R}_i^{GFBC}$ as follows:*

$$\mathfrak{R}_i^{GFBC} \triangleq \bigcup_{\substack{\alpha(.), \beta(.) \\ \varphi(.)}} \begin{Bmatrix} (R_0, R_1, R_2) \in \mathbb{R}_+^3 : \\ R_0 + R_1 \leq \mathbb{E}\left[\psi\left(\frac{|S_1|^2(1-\alpha(E))\varphi(E)}{|S_1|^2\alpha(E)\varphi(E) + 1}\right)\right] \\ R_0 + R_2 \leq \mathbb{E}\left[\psi\left(\frac{|S_2|^2(1-\beta(E))\varphi(E)}{|S_2|^2\beta(E)\varphi(E) + 1}\right)\right] \\ R_0 + R_1 + R_2 \leq \mathbb{E}\left[\psi\left(\frac{|S_1|^2\beta(E)\varphi(E)}{|S_1|^2\alpha(E)\varphi(E) + 1}\right)\right] + \mathbb{E}\left[\psi\left(\frac{|S_2|^2(1-\beta(E))\varphi(E)}{|S_2|^2\beta(E)\varphi(E) + 1}\right)\right] \\ R_0 + R_1 + R_2 \leq \mathbb{E}\left[\psi\left(\frac{|S_1|^2(1-\alpha(E))\varphi(E)}{|S_1|^2\alpha(E)\varphi(E) + 1}\right)\right] + \mathbb{E}\left[\psi\left(\frac{|S_2|^2\alpha(E)\varphi(E)}{|S_2|^2\beta(E)\varphi(E) + 1}\right)\right] \end{Bmatrix}$$

(3)

*where $\varphi(.): \mathcal{E} \to \mathbb{R}_+$ is a power allocation policy function for the transmitter with $\mathbb{E}[\varphi(E)] \leq P$ and also $\alpha(.): \mathcal{E} \to [0,1]$ and $\beta(.): \mathcal{E} \to [0,1]$ are two arbitrary deterministic functions with $\alpha(e) + \beta(e) \leq 1$ for all $e \in \mathcal{E}$. The set $\mathfrak{R}_i^{GFBC}$ constitutes an inner bound on the capacity region of the two-user Gaussian fading BC (1) with common message.*

*Proof of Proposition 1)* The rate region $\mathfrak{R}_i^{GFBC}$ in (3) is derived as a subset of the general rate region $\mathfrak{R}_i^{Marton \to GFBC}$ in (2) by presenting a novel signaling scheme. Assume that $U, V, W$ are independent Gaussian random variables (also independent of the state) with zero mean and unit variance. Define:

$$X \triangleq \sqrt{\varphi(E)}\left(\sqrt{\beta(E)}U + \sqrt{1 - \alpha(E) - \beta(E)}W + \sqrt{\alpha(E)}V\right)$$

(4)

One can readily check that $\mathbb{E}[|X|^2] \leq P$. Now by setting $X, U, V, W$ in $\mathfrak{R}_i^{Marton \to GFBC}$ given by (2), we obtain the achievable rat region (3). Let us interpret the signaling in (4). First, we briefly review the Marton's coding scheme for achieving the rate region (2). See our previous work [26] for a detailed discussion. Consider broadcasting the messages $M_0, M_1, M_2$ to two receivers where the first receiver

Reza K. Farsani, 2013

is required to decode the messages $M_0, M_1$ and the second to decode the messages $M_0, M_2$. Roughly speaking, in the Marton's coding scheme (for a length-$n$ code) each of the messages is split[2] into two parts as:

$$M_1 \triangleq (M_{10}, M_{11}), \qquad M_2 \triangleq (M_{20}, M_{22})$$

Then, the messages $(M_0, M_{10}, M_{20})$ are encoded as common information by a codeword $W^n$ generated based on $P_W$. With respect to each of the sub-messages $M_{11}$ and $M_{22}$, a bin of codewords is randomly generated which are superimposed upon the common information codeword $W^n$: The bin with respect to $M_{11}$ contains the codewords $U^n$ generated based on $P_{U|W}$ and that one for $M_{22}$ contains the codeword $V^n$ generated based on $P_{V|W}$. These bins are then explored against each other to find a jointly typical pair of codewords. Using the mutual covering lemma [25], the sizes of the bins are chosen so large to guarantee the existence of such a jointly typical pair. Superimposed on the designated jointly typical codewords $W^n, U^n, V^n$, the encoder then generates its codewords $X^n$ based on $P_{X|WUV}$, and sends it over the channel. The first receiver decodes the codewords $W^n, U^n$ and the second one decodes $W^n, V^n$, both using a jointly typical decoder. Now let us turn to our signaling scheme in (4). Since we have imposed that the random signals $U, V, W$ are independent, our scheme does not contain any binning scheme. For the case of $\alpha(E) \equiv 0$ (or $V \equiv \emptyset$), our signaling (4) indeed represents a superposition coding wherein the satellite codeword conveys information for the first user. This signaling is useful for the cases where the second receiver is a degraded version of the first one, i.e., $|S_2| < |S_1|$. Similarly, for the case of $\beta(E) \equiv 0$ (or $U \equiv \emptyset$) our signaling represents a superposition coding wherein the satellite codeword conveys information for the second user. This scheme is useful for the cases where the first receiver is a degraded version of the second one, i.e., $|S_1| < |S_2|$. Note that in both schemes, the cloud center codeword conveys information for both users. Therefore, our signaling in (4) contains these two types of superposition coding schemes, simultaneously. This combination strategy is beneficial, because for the fading channel, due to the time-varying nature of the system, in some times the channel is degraded in the sense of $|S_2| < |S_1|$ and in some times is (reversely) degraded in the sense of $|S_1| < |S_2|$. ∎

**Remark 1:** Consider the special case of no side information at the transmitter, i.e., $E \equiv \emptyset$. Our achievability scheme strictly includes the previously proposed one in [2, 27] as a subset. In fact, in the scheme of [2, 27] the transmitter applies superposition coding only in one direction; in other words, the achievable rate region of [2, 27] is derived by setting $\beta(.) \equiv 0$ (or $\alpha(.) \equiv 0$) in our rate region (3). Therefore, unlike our signaling in (4), the proposed scheme of [2, 27] is suitable only for those channels which are uniformly degraded, i.e., the probability of the event $\{|S_2| < |S_1|\}$ is equal to 0 or 1.

We next establish a capacity outer bound for the channel. To this end, we first show that how one can adapt the structure of the UV-outer bound [20] to be applicable for the Gaussian fading channels with stationary state process. This is given in the following lemma.

**Lemma 2)** *Consider the rate region $\mathfrak{R}_o^{UV \to GFBC}$ below:*

$$\mathfrak{R}_o^{UV \to GFBC} \triangleq \bigcup_{\substack{P_{X|E}P_{UV|XS} \\ \mathbb{E}[|X|^2] \leq P}} \begin{Bmatrix} (R_0, R_1, R_2) \in \mathbb{R}_+^3 : \\ R_0 + R_1 \leq I(U; Y_1|\boldsymbol{S}) \\ R_0 + R_2 \leq I(V; Y_2|\boldsymbol{S}) \\ R_0 + R_1 + R_2 \leq I(X; Y_1|V, \boldsymbol{S}) + I(V; Y_2|\boldsymbol{S}) \\ R_0 + R_1 + R_2 \leq I(X; Y_2|U, \boldsymbol{S}) + I(U; Y_1|\boldsymbol{S}) \end{Bmatrix}$$

(5)

*The set $\mathfrak{R}_o^{UV \to GFBC}$ constitutes an outer bound on the capacity region of the two-user Gaussian fading BC (1) with common message.*

*Proof of Lemma 2)* Consider a length-$n$ code with the rate $(R_0, R_1, R_2)$ and vanishing average error probability for the channel. Based on the Fano's inequality we have:

$$H(M_0, M_1|Y_1^n, \boldsymbol{S}^n) \leq n\epsilon_{1,n}$$
$$H(M_0, M_2|Y_2^n, \boldsymbol{S}^n) \leq n\epsilon_{2,n}$$

(6)

where $\epsilon_{1,n}, \epsilon_{2,n} \to 0$ as $n \to \infty$. Define new auxiliary random variables $U_t$ and $V_t$ as follows:

---

[2] Alternatively, one may ignore the message splitting and instead at the last step enlarge the resultant rate region using the fact that if the triple $(R_0, R_1, R_2)$ is achievable for the two-user BC, then $(R_0 - \tau_1 - \tau_2, R_1 + \tau_1, R_2 + \tau_2)$ is also achievable. However, to interpret our signaling in (4), the message-splitting approach is more useful, because by nullifying either $U$ or $V$, the scheme directly is reduced to the superposition coding for broadcasting the common and *both private messages*.

$$\begin{cases} U_t \triangleq \left(Y_{1,t+1}^n, Y_2^{t-1}, M_0, M_1, \mathbf{S}^{t-1}, \mathbf{S}_{t+1}^n\right) \\ V_t \triangleq \left(Y_{1,t+1}^n, Y_2^{t-1}, M_0, M_2, \mathbf{S}^{t-1}, \mathbf{S}_{t+1}^n\right) \end{cases}, \quad t = 1, \dots, n \tag{7}$$

Therefore, we have:

$$\begin{aligned}
n(R_0 + R_1) &\leq I(M_0, M_1; Y_1^n, \mathbf{S}^n) + n\epsilon_{1,n} \\
&\stackrel{(a)}{=} I(M_0, M_1; Y_1^n | \mathbf{S}^n) + n\epsilon_{1,n} \\
&= \sum_{t=1}^n I(M_0, M_1; Y_{1,t} | Y_{1,t+1}^n, \mathbf{S}^n) + n\epsilon_{1,n} \\
&\leq \sum_{t=1}^n I(Y_{1,t+1}^n, Y_2^{t-1}, M_0, M_1, \mathbf{S}^{t-1}, \mathbf{S}_{t+1}^n; Y_{1,t} | \mathbf{S}_t) + n\epsilon_{1,n} = \sum_{t=1}^n I(U_t; Y_{1,t} | \mathbf{S}_t) + n\epsilon_{1,n}
\end{aligned} \tag{8}$$

where equality (a) holds because $M_0, M_1, M_2$ and $\mathbf{S}^n$ are independent. Also,

$$\begin{aligned}
n(R_0 + R_1 + R_2) &\leq I(M_2; Y_2^n, \mathbf{S}^n) + I(M_0, M_1; Y_1^n, \mathbf{S}^n) + n(\epsilon_{1,n} + \epsilon_{2,n}) \\
&\leq I(M_2; Y_2^n | M_0, M_1, \mathbf{S}^n) + I(M_0, M_1; Y_1^n | \mathbf{S}^n) + n(\epsilon_{1,n} + \epsilon_{2,n}) \\
&= \sum_{t=1}^n I(M_2; Y_{2,t} | Y_2^{t-1}, M_0, M_1, \mathbf{S}^n) + \sum_{t=1}^n I(M_0, M_1; Y_{1,t} | Y_{1,t+1}^n, \mathbf{S}^n) + n(\epsilon_{1,n} + \epsilon_{2,n}) \\
&\leq \sum_{t=1}^n I(M_2, Y_{1,t+1}^n; Y_{2,t} | Y_2^{t-1}, M_0, M_1, \mathbf{S}^n) + \sum_{t=1}^n I(Y_2^{t-1}, M_0, M_1; Y_{1,t} | Y_{1,t+1}^n, \mathbf{S}^n) \\
&\quad - \sum_{t=1}^n I(Y_2^{t-1}; Y_{1,t} | Y_{1,t+1}^n, M_0, M_1, \mathbf{S}^n) + n(\epsilon_{1,n} + \epsilon_{2,n}) \\
&= \sum_{t=1}^n I(M_2; Y_{2,t} | Y_{1,t+1}^n, Y_2^{t-1}, M_0, M_1, \mathbf{S}^n) + \sum_{t=1}^n I(Y_2^{t-1}, M_0, M_1; Y_{1,t} | Y_{1,t+1}^n, \mathbf{S}^n) \\
&\quad + \sum_{t=1}^n I(Y_{1,t+1}^n; Y_{2,t} | Y_2^{t-1}, M_0, M_1, \mathbf{S}^n) - \sum_{t=1}^n I(Y_2^{t-1}; Y_{1,t} | Y_{1,t+1}^n, M_0, M_1, \mathbf{S}^n) + n(\epsilon_{1,n} + \epsilon_{2,n}) \\
&\stackrel{(a)}{=} \sum_{t=1}^n I(M_2; Y_{2,t} | Y_{1,t+1}^n, Y_2^{t-1}, M_0, M_1, \mathbf{S}^n) + \sum_{t=1}^n I(Y_2^{t-1}, M_0, M_1; Y_{1,t} | Y_{1,t+1}^n, \mathbf{S}^n) + n(\epsilon_{1,n} + \epsilon_{2,n}) \\
&\stackrel{(b)}{\leq} \sum_{t=1}^n I(M_2; Y_{2,t} | Y_{1,t+1}^n, Y_2^{t-1}, M_0, M_1, \mathbf{S}^{t-1}, \mathbf{S}_{t+1}^n, \mathbf{S}_t) + \sum_{t=1}^n I(Y_{1,t+1}^n, Y_2^{t-1}, M_0, M_1, \mathbf{S}^{t-1}, \mathbf{S}_{t+1}^n; Y_{1,t} | \mathbf{S}_t) + n(\epsilon_{1,n} + \epsilon_{2,n}) \\
&\stackrel{(c)}{=} \sum_{t=1}^n I(X_t, M_2; Y_{2,t} | U_t, \mathbf{S}_t) + \sum_{t=1}^n I(U_t; Y_{1,t} | \mathbf{S}_t) + n(\epsilon_{1,n} + \epsilon_{2,n}) \\
&\stackrel{(d)}{=} \sum_{t=1}^n I(X_t; Y_{2,t} | U_t, \mathbf{S}_t) + \sum_{t=1}^n I(U_t; Y_{1,t} | \mathbf{S}_t) + n(\epsilon_{1,n} + \epsilon_{2,n})
\end{aligned} \tag{9}$$

where equality (a) is due to the Csiszar-Korner identity, inequality (b) holds because conditioning does not increase the entropy, equality (c) holds because $X_t$ is a deterministic function of $(M_0, M_1, M_2, E_t = \xi(\mathbf{S}_t))$, and equality (d) holds because $M_2, U_t \to X_t, \mathbf{S}_t \to Y_{2,t}$ forms a Markov chain. By following a rather similar procedure, one can derive:

$$\begin{aligned}
n(R_0 + R_2) &\leq \sum_{t=1}^n I(V_t; Y_{2,t} | \mathbf{S}_t) + n\epsilon_{2,n} \\
n(R_0 + R_1 + R_2) &\leq \sum_{t=1}^n I(X_t; Y_{1,t} | V_t, \mathbf{S}_t) + \sum_{t=1}^n I(V_t; Y_{2,t} | \mathbf{S}_t) + n(\epsilon_{1,n} + \epsilon_{2,n})
\end{aligned} \tag{10}$$

Then, we introduce a time-sharing random variable $Q$ uniformly distributed over the set $\{1, \dots, n\}$ and independent of other RVs. Define:

$$U \triangleq (U_Q, Q), \quad V \triangleq (V_Q, Q), \quad X \triangleq X_Q, \quad Y_1 \triangleq Y_{1,Q}, \quad Y_2 \triangleq Y_{2,Q}, \quad \mathbf{S} \triangleq \mathbf{S}_Q, \quad Z \triangleq Z_Q \tag{11}$$

Note that $\mathbf{S}$ and $Z$ are indeed independent of $Q$, because the state process of the channel is stationary and the noise process is i.i.d. Now, based on (11) we can write:

$$\begin{aligned}
R_0 + R_1 &\leq \sum_{t=1}^n \frac{1}{n} I(U_t; Y_{1,t} | \mathbf{S}_t) + \epsilon_{1,n} = I(U_Q; Y_{1,Q} | \mathbf{S}_Q, Q) + \epsilon_{1,n} \\
&\leq I(U_Q, Q; Y_{1,Q} | \mathbf{S}_Q) + \epsilon_{1,n} = I(U; Y_1 | \mathbf{S}) + \epsilon_{1,n}
\end{aligned} \tag{12}$$

Other summations included in the bounds (9)-(10) can also be expressed in such a compact form, similarly. Then, by letting $n \to \infty$, we obtain the constraints in (5). Moreover, we have:

$$\mathbb{E}[|X|^2] = \frac{1}{n}\mathbb{E}\left[\sum_{t=1}^{n}|X_t|^2\right] \leq P \tag{13}$$

The proof of Lemma 1 is thus complete. ∎

***Remarks 2:***

1. Consider the definitions of the random variables $U_t$ and $V_t$, where $t = 1, ..., n$, in (7). It is should be noted that these random variables both are correlated to the state of the channel, i.e., $\boldsymbol{S}_t$, because the state process is stationary. This point is of great importance when optimizing the rate region (5). Nevertheless, if we impose that the state process of the channel is i.i.d., then $U_t$ and $V_t$ both are independent of $\boldsymbol{S}_t$. Therefore, for the channels with i.i.d. state process when optimizing the rate region (5), we can restrict our attention to the space of all random variables $U$ and $V$ which are independent of $\boldsymbol{S}$.

2. Instead of our style proof for the derivation of the outer bound in (5), one may think that it is initially possible to replace $Y_1$ by $(Y_1, \boldsymbol{S})$ and $Y_2$ by $(Y_2, \boldsymbol{S})$ in the UV-outer bound of [20] and directly obtain the rate region (5). Let us describe why this naive idea fails. By the latter substitution in the UV-outer bound of [20], for example for the sum-rate we obtain:

$$R_1 + R_2 \leq I(X; Y_2, \boldsymbol{S}|U) + I(U; Y_1, \boldsymbol{S})$$

Now note that when the transmitter has access to side information, $X$ depends on the state $\boldsymbol{S}$. Even when the transmitter has no side information, since the state process is stationary, according to the previous remark, the random variables $U$ and $V$ both are correlated with $\boldsymbol{S}$. Therefore, we have:

$$I(X; Y_2, \boldsymbol{S}|U) + I(U; Y_1, \boldsymbol{S}) \gtreqless I(X; Y_2|U, \boldsymbol{S}) + I(U; Y_1|\boldsymbol{S})$$

In other words, we cannot derive the outer bound (5). Only for the case where the state process is i.i.d. and the transmitter has no side information, the idea of replacing $Y_1$ by $(Y_1, \boldsymbol{S})$ and $Y_2$ by $(Y_2, \boldsymbol{S})$ in the UV-outer bound of [20] does work.

Then, we explicitly evaluate the outer bound (5) by a novel approach, as given below.

***Theorem 1)*** *Consider the two-user Gaussian fading BC (1) with common message. Define the rate region $\mathfrak{R}_o^{GFBC}$ as follows:*

$$\mathfrak{R}_o^{GFBC} \triangleq \bigcup_{\substack{\alpha(.),\beta(.) \\ \varphi(.)}} \left\{ \begin{array}{l} (R_0, R_1, R_2) \in \mathbb{R}_+^3: \\[4pt] R_0 + R_1 \leq \mathbb{E}\left[\psi\left(\frac{|S_1|^2(1-\alpha(\boldsymbol{S}))\varphi(E)\mathbb{1}(|S_1|<|S_2|)}{|S_1|^2\alpha(\boldsymbol{S})\varphi(E)+1}\right)\right] + \mathbb{E}\left[\psi\left(|S_1|^2\varphi(E)\mathbb{1}(|S_1| \geq |S_2|)\right)\right], \\[8pt] R_0 + R_2 \leq \mathbb{E}\left[\psi\left(\frac{|S_2|^2(1-\beta(\boldsymbol{S}))\varphi(E)\mathbb{1}(|S_2|<|S_1|)}{|S_2|^2\beta(\boldsymbol{S})\varphi(E)+1}\right)\right] + \mathbb{E}\left[\psi\left(|S_2|^2\varphi(E)\mathbb{1}(|S_2| \geq |S_1|)\right)\right], \\[8pt] R_0 + R_1 + R_2 \leq \mathbb{E}\left[\psi\left(|S_1|^2\beta(\boldsymbol{S})\varphi(E)\mathbb{1}(|S_2|<|S_1|)\right) + \psi\left(\frac{|S_2|^2(1-\beta(\boldsymbol{S}))\varphi(E)\mathbb{1}(|S_2|<|S_1|)}{|S_2|^2\beta(\boldsymbol{S})\varphi(E)+1}\right)\right] \\[4pt] \qquad\qquad\qquad +\mathbb{E}\left[\psi\left(|S_2|^2\varphi(E)\mathbb{1}(|S_2| \geq |S_1|)\right)\right], \\[8pt] R_0 + R_1 + R_2 \leq \mathbb{E}\left[\psi\left(|S_2|^2\alpha(\boldsymbol{S})\varphi(E)\mathbb{1}(|S_1|<|S_2|)\right) + \psi\left(\frac{|S_1|^2(1-\alpha(\boldsymbol{S}))\varphi(E)\mathbb{1}(|S_1|<|S_2|)}{|S_1|^2\alpha(\boldsymbol{S})\varphi(E)+1}\right)\right] \\[4pt] \qquad\qquad\qquad +\mathbb{E}\left[\psi\left(|S_1|^2\varphi(E)\mathbb{1}(|S_1| \geq |S_2|)\right)\right] \end{array} \right\} \tag{14}$$

*where $\alpha(.): \mathbb{C}^2 \to [0,1]$ and $\beta(.): \mathbb{C}^2 \to [0,1]$ are arbitrary deterministic functions; also, $\varphi(.): \mathcal{E} \to \mathbb{R}_+$ with $\mathbb{E}[\varphi(E)] \leq P$ denotes the power allocation policy for the transmitter. The set $\mathfrak{R}_o^{GFBC}$ constitutes an outer bound on the capacity region.*

*Proof of Theorem 1)* To derive the outer bound (14), we present novel arguments to optimize the rate region $\mathfrak{R}_o^{UV \to GFBC}$ in (5) for all joint PDFs $P_{X|E}(x|e)P_{UV|XS}(u,v|x,s)$ with $\mathbb{E}[|X|^2] \leq P$. Let us first point out a previous effort to solve a similar optimization problem, although for the special case with no side information at the transmitter and also i.i.d. state process. Specifically, consider the following constraints of the UV-outer bound $\mathfrak{R}_o^{UV \to GFBC}$ in (5), (for the case of $R_0 = 0$):

$$R_1 \leq I(U; Y_1 | \boldsymbol{S})$$
$$R_1 + R_2 \leq I(X; Y_2 | U, \boldsymbol{S}) + I(U; Y_1 | \boldsymbol{S})$$
(15)

These constraints should be optimized for all joint PDFs $P_{XU}(x, u)$. The authors in [2] (see also [27-28]) examined the procedure below for solving the problem. We have:

$$I(U; Y_1 | \boldsymbol{S}) = H(Y_1 | \boldsymbol{S}) - H(Y_1 | U, \boldsymbol{S}) \stackrel{(a)}{\leq} \mathbb{E}[\log \pi e(|S_2|^2 P + 1)] - H(Y_1 | U, \boldsymbol{S})$$
$$I(X; Y_2 | U, \boldsymbol{S}) = H(Y_2 | U, \boldsymbol{S}) - H(Y_2 | X, U, \boldsymbol{S}) = H(Y_2 | U, \boldsymbol{S}) - \log \pi e$$
(16)

where inequality (a) is due to the "Gaussian maximizes the entropy" principle. Now consider the term $H(Y_2 | U, \boldsymbol{S})$. We have:

$$\log \pi e = H(Z_2) \leq H(Y_2 | U, \boldsymbol{S}) = H(S_2 X + Z_2 | U, \boldsymbol{S}) \leq H(S_2 X + Z_2 | \boldsymbol{S}) \leq \mathbb{E}[\log \pi e(|S_2|^2 P + 1)]$$
(17)

The authors [2, 27-28] then argued that (17) implies that there exists $\alpha$ belonging to the interval $[0,1]$ such that:

$$H(Y_2 | U, \boldsymbol{S}) = \mathbb{E}[\log \pi e(\alpha |S_2|^2 P + 1)]$$
(18)

However, as indicted in [28], by this procedure the use of EPI fails for bounding the term $H(Y_1|U, \boldsymbol{S})$ in (16), even for the case where the channel is uniformly degraded. The above arguments indeed are reminiscent of the proof of Bergmans [29] for the converse of Gaussian non-fading BC. But it seems this approach is not applicable for the fading channel. It has been also remarked in [6] that the conventional EPI is not directly applicable for the fading BC. Nevertheless, in what follows we present novel arguments based on which the outer bound $\mathfrak{R}_o^{UV \to GFBC}$ in (5) can still be evaluated using the EPI, not only for the special case with no CSIT and i.i.d. state process but also for the general channel with any arbitrary CSIT and stationary state process. Roughly speaking, by taking integral over the state $\boldsymbol{S}$ we evaluate the right side of the constraints in (16) for each state $\boldsymbol{S} = \boldsymbol{s}$. Accordingly, $\alpha$ in equations such as (17) is no more a constant parameter belonging to $[0,1]$; instead, it would be a deterministic function of the state with the range of $[0,1]$. Moreover, we evaluate the sum of the two mutual information functions in the second constraint of (15), totally, unlike the procedure of (16)-(18) in which each mutual information function is evaluated separately (the reason for this step will be clarified later). By this approach, we can apply the EPI to optimize the constraints.

Note that it is only required to evaluate $1^{th}$ and $4^{th}$ constraints of $\mathfrak{R}_o^{UV \to GFBC}$ in (5) because the two other constraints can be evaluated symmetrically. Fix a joint PDF $P_{X|E}(x|e) P_{UV|XS}(u, v|x, \boldsymbol{s})$ with $\mathbb{E}[|X|^2] \leq P$. Define the deterministic function $\varphi(.)$ as follows:

$$\varphi(.): \mathcal{E} \to \mathbb{R}_+, \qquad \varphi(e) \triangleq \mathbb{E}[|X|^2 | E = e]$$
(19)

Thereby, we have: $\mathbb{E}[\varphi(E)] \leq P$. For $1^{th}$ and $4^{th}$ constraints of $\mathfrak{R}_o^{UV \to GFBC}$ in (5), one can write:

$$I(U; Y_1|\boldsymbol{S}) = \int_{|S_1|<|S_2|} P_{\boldsymbol{S}}(\boldsymbol{s}) I(U; Y_1|\boldsymbol{s}) + \int_{|S_1|\geq|S_2|} P_{\boldsymbol{S}}(\boldsymbol{s}) I(U; Y_1|\boldsymbol{s})$$
(20)

$$I(X; Y_2|U, \boldsymbol{S}) + I(U; Y_1|\boldsymbol{S}) = \int_{|S_1|<|S_2|} P_{\boldsymbol{S}}(\boldsymbol{s}) \big(I(X; Y_2|U, \boldsymbol{s}) + I(U; Y_1|\boldsymbol{s})\big) + \int_{|S_1|\geq|S_2|} P_{\boldsymbol{S}}(\boldsymbol{s}) \big(I(X; Y_2|U, \boldsymbol{s}) + I(U; Y_1|\boldsymbol{s})\big)$$
(21)

Consider the first integrals in (20) and (21). Let $\boldsymbol{s} \in \{|S_1| < |S_2|\}$. Let also $\tilde{Z}_1$ be a Gaussian virtual noise, independent of $Z_1$ and $Z_2$, with zero mean and unit variance. We have:

$$I(U;Y_1|\boldsymbol{s}) = I\left(U; \frac{s_1}{s_2}Y_2 + \sqrt{1 - \left|\frac{s_1}{s_2}\right|^2}\tilde{Z}_1 \middle| \boldsymbol{s}\right)$$

$$= H\left(s_1 X + \frac{s_1}{s_2}Z_2 + \sqrt{1 - \left|\frac{s_1}{s_2}\right|^2}\tilde{Z}_1 \middle| \boldsymbol{s}\right) - H\left(\frac{s_1}{s_2}Y_2 + \sqrt{1 - \left|\frac{s_1}{s_2}\right|^2}\tilde{Z}_1 \middle| U, \boldsymbol{s}\right)$$

$$\overset{(a)}{\leq} \log \pi e(|s_1|^2 \mathbb{E}[|X|^2|E = e] + 1) - H\left(\frac{s_1}{s_2}Y_2 + \sqrt{1 - \left|\frac{s_1}{s_2}\right|^2}\tilde{Z}_1 \middle| U, \boldsymbol{s}\right)$$

$$= \log \pi e(|s_1|^2 \varphi(e) + 1) - H\left(\frac{s_1}{s_2}Y_2 + \sqrt{1 - \left|\frac{s_1}{s_2}\right|^2}\tilde{Z}_1 \middle| U, \boldsymbol{s}\right)$$

(22)

where (a) is due to the "Gaussian maximizes the entropy" principle. Also,

$$I(X;Y_2|U,\boldsymbol{s}) + I(U;Y_1|\boldsymbol{s}) = H(Y_2|U,\boldsymbol{s}) - H(s_2 X + Z_2|X,U,\boldsymbol{s}) + I(U;Y_1|\boldsymbol{s})$$

$$= H(Y_2|U,\boldsymbol{s}) - H(s_2 X + Z_2|X,U,\boldsymbol{s}) + H(s_1 X + Z_1|\boldsymbol{s}) - H(Y_1|U,\boldsymbol{s})$$

$$\leq H(Y_2|U,\boldsymbol{s}) - \log \pi e + \log \pi e(|s_1|^2 \varphi(e) + 1) - H\left(\frac{s_1}{s_2}Y_2 + \sqrt{1 - \left|\frac{s_1}{s_2}\right|^2}\tilde{Z}_1 \middle| U, \boldsymbol{s}\right)$$

(23)

Now let evaluate the term $H(Y_2|U,\boldsymbol{s}) = H(s_2 X + Z_2|U,\boldsymbol{s})$ in (23). We have:

$$\log \pi e = H(Z_2) \leq H(Y_2|U,\boldsymbol{s}) = H(s_2 X + Z_2|U,\boldsymbol{s}) \leq H(s_2 X + Z_2|\boldsymbol{s}) \leq \log \pi e(|s_2|^2 \varphi(e) + 1)$$

(24)

The two sides of (24) imply that there exist $0 \leq \alpha(\boldsymbol{s}) \leq 1$ such that:

$$H(Y_2|U,\boldsymbol{s}) = H(s_2 X + Z_2|U,\boldsymbol{s}) = \log \pi e(|s_2|^2 \alpha(\boldsymbol{s})\varphi(e) + 1)$$

(25)

Then, we bound the term $H(Y_1|U,\boldsymbol{s}) = H\left(\frac{s_1}{s_2}Y_2 + \sqrt{1 - \left|\frac{s_1}{s_2}\right|^2}\tilde{Z}_1 \middle| U, \boldsymbol{s}\right)$ in (22) and (23) as follows:

$$H(Y_1|U,\boldsymbol{s}) = H\left(\frac{s_1}{s_2}Y_2 + \sqrt{1 - \left|\frac{s_1}{s_2}\right|^2}\tilde{Z}_1 \middle| U, \boldsymbol{s}\right) \overset{(a)}{\geq} \log\left(2^{H\left(\frac{s_1}{s_2}Y_2 \middle| U,\boldsymbol{s}\right)} + 2^{H\left(\sqrt{1-\left|\frac{s_1}{s_2}\right|^2}\tilde{Z}_1 \middle| U,\boldsymbol{s}\right)}\right)$$

$$= \log\left(\frac{|s_1|^2}{|s_2|^2} 2^{H(Y_2|U,\boldsymbol{s})} + \pi e\left(1 - \frac{|s_1|^2}{|s_2|^2}\right)\right)$$

$$\overset{(b)}{=} \log \pi e(|s_1|^2 \alpha(\boldsymbol{s})\varphi(e) + 1)$$

(26)

where (a) is due to the EPI and (b) is derived by (25). Therefore, from (22), (23), (25) and (26) we obtain:

$$\int_{|s_1|<|s_2|} P_{\boldsymbol{S}}(\boldsymbol{s}) I(U;Y_1|\boldsymbol{s}) \leq \int_{|s_1|<|s_2|} P_{\boldsymbol{S}}(\boldsymbol{s}) \psi\left(\frac{|s_1|^2(1-\alpha(\boldsymbol{s}))\varphi(e)}{|s_1|^2 \alpha(\boldsymbol{s})\varphi(e) + 1}\right) = \mathbb{E}\left[\psi\left(\frac{|S_1|^2(1-\alpha(\boldsymbol{S}))\varphi(E)\mathbb{1}(|S_1|<|S_2|)}{|S_1|^2 \alpha(\boldsymbol{S})\varphi(E) + 1}\right)\right]$$

(27)

$$\int_{|s_1|<|s_2|} P_{\boldsymbol{S}}(\boldsymbol{s})\big(I(X;Y_2|U,\boldsymbol{s}) + I(U;Y_1|\boldsymbol{s})\big) \leq \int_{|s_1|<|s_2|} P_{\boldsymbol{S}}(\boldsymbol{s})\left(\psi(|s_2|^2 \alpha(\boldsymbol{s})\varphi(e)) + \psi\left(\frac{|s_1|^2(1-\alpha(\boldsymbol{s}))\varphi(e)}{|s_1|^2 \alpha(\boldsymbol{s})\varphi(e) + 1}\right)\right)$$

$$= \mathbb{E}\left[\psi(|S_2|^2 \alpha(\boldsymbol{S})\varphi(E)\mathbb{1}(|S_1|<|S_2|)) + \psi\left(\frac{|S_1|^2(1-\alpha(\boldsymbol{S}))\varphi(E)\mathbb{1}(|S_1|<|S_2|)}{|S_1|^2 \alpha(\boldsymbol{S})\varphi(E) + 1}\right)\right]$$

(28)

Next, consider the second integrals in (20) and (21). We have:

$$\int_{|s_1|\geq|s_2|} P_S(s) I(U;Y_1|s) \leq \int_{|s_1|\geq|s_2|} P_S(s) I(X;Y_1|s) \leq \mathbb{E}\big[\psi\big(|S_1|^2 \varphi(E)\mathbb{1}(|S_1|\geq|S_2|)\big)\big]$$

(29)

Also,

$$\int_{|s_1|\geq|s_2|} P_S(s)\big(I(X;Y_2|U,s) + I(U;Y_1|s)\big) \stackrel{(a)}{\leq} \int_{|s_1|\geq|s_2|} P_S(s)\big(I(X;Y_1|U,s) + I(U;Y_1|s)\big)$$

$$= \int_{|s_1|\geq|s_2|} P_S(s) I(X;Y_1|s) \leq \mathbb{E}\big[\psi\big(|S_1|^2 \varphi(E)\mathbb{1}(|S_1|\geq|S_2|)\big)\big]$$

(30)

where inequality (a) holds because for $|s_1|\geq|s_2|$, the receiver $Y_2$ is a degraded version of $Y_1$. Note that in the last step, i.e., equation (30), it is critical to totally optimize the sum expression $I(X;Y_2|U,s) + I(U;Y_1|s)$ because if we would independently optimize each of the mutual information functions $I(X;Y_2|U,s)$ and $I(U;Y_1|s)$ with $s \in \{|S_1|\geq|S_2|\}$, we get $I(X;Y_2|s)$ and $I(X;Y_1|s)$, respectively. The fact is that in (15) and (16) the auxiliary random variable $U$ is enhanced to $X$. By substituting (27)-(30) in (20) and (21), we derive the desired constraints in (14). The proof of Theorem 1 is complete. ∎

**Remark 3:** The outer bound $\mathfrak{R}_o^{GFBC}$ given in (14) is uniformly applicable for all situations with arbitrary fading statistics and arbitrary amount of state information at the transmitter.

We next prove that for the channel with perfect CSIT, i.e., $E \equiv S$, the derived inner and outer bounds coincide. This result is given in the following theorem.

**Theorem 2)** *Consider the two-user Gaussian fading BC (1) with common message wherein the transmitter knows the state perfectly, i.e., $E \equiv S$. The inner bound $\mathfrak{R}_i^{GFBC}$ in (3) and the outer bound $\mathfrak{R}_o^{GFBC}$ in (14) coincide and result to the capacity region.*

*Proof of Theorem 2)* Let $\alpha^*(.): \mathbb{C}^2 \to [0,1]$ and $\beta^*(.): \mathbb{C}^2 \to [0,1]$ be two arbitrary deterministic functions. Define the deterministic functions $\alpha(.): \mathbb{C}^2 \to [0,1]$ and $\beta(.): \mathbb{C}^2 \to [0,1]$ as follows:

$$\alpha(S) \triangleq \begin{cases} \alpha^*(S) & \text{if } |S_1| < |S_2| \\ 0 & \text{if } |S_1| \geq |S_2| \end{cases}, \quad \beta(S) \triangleq \begin{cases} 0 & \text{if } |S_1| < |S_2| \\ \beta^*(S) & \text{if } |S_1| \geq |S_2| \end{cases}$$

(31)

Thereby, we have $\alpha(s) + \beta(s) \leq 1$ for all $s \in \mathbb{C}^2$. Now by substituting $\alpha(.)$ and $\beta(.)$ in the achievable rate region $\mathfrak{R}_i^{GFBC}$ in (3), one can see that it is equal to the rate region $\mathfrak{R}_o^{GFBC}$ in (14) if it is evaluated by $\alpha^*(S)$ and $\beta^*(S)$. The derivation of their equivalence is indeed interesting. ∎

*Remarks 4:*

1. Let us discuss the special case of the channel without common message, i.e., $R_0 = 0$, where the transmitter knows the state perfectly, i.e., $E \equiv S$. Consider the following rate region:

$$\bigcup_{\substack{\alpha(.),\beta(.) \\ \varphi(.)}} \left\{ \begin{aligned} &(R_1, R_2) \in \mathbb{R}_+^2: \\ &R_1 \leq \mathbb{E}\left[\psi\left(\frac{|S_1|^2(1-\alpha(S))\varphi(S)\mathbb{1}(|S_1|<|S_2|)}{|S_1|^2\alpha(S)\varphi(S)+1}\right)\right] + \mathbb{E}\big[\psi\big(|S_1|^2\beta(S)\varphi(S)\mathbb{1}(|S_1|\geq|S_2|)\big)\big] \\ &R_2 \leq \mathbb{E}\left[\psi\left(\frac{|S_2|^2(1-\beta(S))\varphi(S)\mathbb{1}(|S_2|<|S_1|)}{|S_2|^2\beta(S)\varphi(S)+1}\right)\right] + \mathbb{E}\big[\psi\big(|S_2|^2\alpha(S)\varphi(S)\mathbb{1}(|S_2|\geq|S_1|)\big)\big] \end{aligned} \right\}$$

(32)

where $\alpha(.): \mathbb{C}^2 \to [0,1]$, $\beta(.): \mathbb{C}^2 \to [0,1]$, and $\varphi(.): \mathbb{C}^2 \to \mathbb{R}_+$ with $\mathbb{E}[\varphi(S)] \leq P$ are arbitrary deterministic functions. One can show that the rate region (32) is a subset of (14). Moreover, every vertex of the convex hull of the rate region (14) is inside (32). Therefore, the two rate regions are equivalent. Also, note that in the rate region (32) without loss of generality one may impose $\alpha(.) \equiv 1 - \beta(.)$. This rate region was previously derived in [1] as the ergodic capacity of the BC with perfect CSIT. Thus, Theorem 2 establishes a new and more concise proof for the problem. We remark that in [1] the authors showed that the channel with perfect CSIT is decomposed into parallel sub-channels and build their result based on [30]. By our approach, in addition to establishing the capacity region for the channel with perfect CSIT, a useful capacity outer bound is also derived for other cases (the channels with any arbitrary amount of CSIT).

2. Let us review the signaling (4) and (31) that achieve the capacity region for the channel with perfect CSIT. Based on this scheme, the fading channel is divided in two phases: In one phase the channel is degraded in the sense of $|S_1| \geq |S_2|$; in this case the transmitter applies a superposition signaling where the satellite signal is designated for the first user. The portion of power allocated to the cloud center signal is $(1 - \beta(S))\varphi(S)$ and that allocated to the satellite is $\beta(S)\varphi(S)$. In the other phase, the channel is reversely degraded in the sense of $|S_2| \geq |S_1|$; in this phase the transmitter applies a superposition signaling where the satellite signal is designated for the second user. For this case, the portion of power allocated to the cloud center signal is $(1 - \alpha(S))\varphi(S)$ and that allocated to the satellite is $\alpha(S)\varphi(S)$. The availability of the state at the transmitter affects (improves) the capacity from two viewpoints: 1- As the transmitter knows the degradedness ordering of the channel, it can decide that to which of the users the satellite signal should be sent, 2- The transmitter manages the portion of power which should be allocated to each of the cloud center and the satellite signals based on the state information.

For the general case with partial side information at the transmitter, the inner bound (3) and the outer bound (14) may not coincide. The main reason is that the functions $\alpha(.)$ and $\beta(.)$ in the inner bound (3) depend on the side information $E$ while they depend on the state $S$ for the outer bound (14). Nevertheless, one may still explore for the special cases where these bounds coincide or at least have the same maximum sum-rate. We present an instance in the following theorem.

**Theorem 3)** *Consider the two-user Gaussian fading BC (1) with common message. Let the side information available at the transmitter be $E \equiv (E^*, E^\times)$ where $E^* \equiv \mathbb{1}(|S_1| < |S_2|)$ and $E^\times$ is given by an arbitrary deterministic function of the state $S$. The sum-rate capacity is given below:*

$$\max_{\varphi(.): \mathbb{E}[\varphi(E)] \leq P} \left( \mathbb{E}\left[\psi\left(|S_1|^2 \varphi(E) \mathbb{1}(|S_1| \geq |S_2|)\right)\right] + \mathbb{E}\left[\psi\left(|S_2|^2 \varphi(E) \mathbb{1}(|S_2| \geq |S_1|)\right)\right] \right) \tag{33}$$

*Proof of Theorem 3)* The achievability is derived from the inner bound (3) by setting:

$$\alpha(E) \triangleq \begin{cases} 1 & \text{if } E^* = 1 \\ 0 & \text{if } E^* = 0 \end{cases}, \quad \beta(E) \triangleq \begin{cases} 0 & \text{if } E^* = 1 \\ 1 & \text{if } E^* = 0 \end{cases} \tag{34}$$

For the converse part we make use of the outer bound (14). For the sum-rate we have:

$$R_0 + R_1 + R_2 \leq \mathbb{E}\left[\psi\left(|S_1|^2 \beta(S)\varphi(E)\mathbb{1}(|S_2| < |S_1|)\right) + \psi\left(\frac{|S_2|^2(1-\beta(S))\varphi(E)\mathbb{1}(|S_2| < |S_1|)}{|S_2|^2\beta(S)\varphi(E) + 1}\right)\right] + \mathbb{E}\left[\psi\left(|S_2|^2\varphi(E)\mathbb{1}(|S_2| \geq |S_1|)\right)\right]$$

$$\stackrel{(a)}{\leq} \mathbb{E}\left[\psi\left(|S_1|^2\varphi(E)\mathbb{1}(|S_2| < |S_1|)\right)\right] + \mathbb{E}\left[\psi\left(|S_2|^2\varphi(E)\mathbb{1}(|S_2| \geq |S_1|)\right)\right] \tag{35}$$

where the inequality (a) holds because when $|S_2| < |S_1|$, the following expression:

$$\psi\left(|S_1|^2\beta(S)\varphi(E)\right) + \psi\left(\frac{|S_2|^2(1-\beta(S))\varphi(E)}{|S_2|^2\beta(S)\varphi(E) + 1}\right) \tag{36}$$

is monotonically increasing in terms of $\beta(S)$. The proof is thus complete. ∎

One of important scenarios from the viewpoint of practical interests is the fading channel with no CSIT, i.e., $E \equiv \emptyset$. As discussed in introduction, this case has been studied in several papers [2-6], however, the capacity region is still unknown. By setting $E \equiv \emptyset$ and hence $\varphi(.) \equiv P$ in the rate regions (3) and (14), it is clear that inner and outer bounds are derived for this channel. These bounds have similar structures; however, they do not coincide in general because $\alpha(.)$ and $\beta(.)$ in the outer bound depends on the state $\boldsymbol{S} = (S_1, S_2)$.

In the following theorem, we show that if the state process is i.i.d., a potentially tighter outer bound than that given by (14) can also be established.

**Theorem 4)** Consider the two-user Gaussian fading BC (1) with common message and i.i.d. state process. In this case, in the outer bound $\mathfrak{R}_o^{GFBC}$ in (14) one can restrict the functions $\alpha(.)$ and $\beta(.)$ to depend only on $(S_2, E)$ and $(S_1, E)$, respectively, and not on the whole of the state $\boldsymbol{S} = (S_1, S_2)$. Specially, if there is no side information at the transmitter, i.e., $E \equiv \emptyset$, then $\alpha(.)$ and $\beta(.)$ are decreasing functions[3] of $S_2$ and $S_1$, respectively.

*Proof of Theorem 4)* Consider the left side of the equation (25) based on which $\alpha(.)$ was defined. As discussed in Remark 2, when the state process is i.i.d, in the UV-outer bound (5) one can impose that the auxiliary random variables $U$ and $V$ are both independent of the state $\boldsymbol{S}$. Also, note that according to the definition (7), the input signal $X$ is indeed a deterministic function of $(U, V, E)$. Therefore, one can easily show that for the case of i.i.d. state process the Markov relation $\boldsymbol{S} \to E \to (UVX)$ holds. Consequently, we have the following equality:

$$H(Y_2|U, \boldsymbol{s}) = H(s_2 X + Z_2|U, \boldsymbol{s}) = H(s_2 X + Z_2|U, e) \tag{37}$$

Now, considering the two sides of the inequalities (24), we deduce that with respect to each $(s_2, e)$ there exist $\alpha(s_2, e)$ such that:

$$H(Y_2|U, \boldsymbol{s}) = H(s_2 X + Z_2|U, e) = \log \pi e(|s_2|^2 \alpha(s_2, e)\varphi(e) + 1) \tag{38}$$

In other words, the function $\alpha(.)$ depends only on $(S_2, E)$. Next assume that there is no side information at the transmitter, i.e., $E \equiv \emptyset$. In this case, $X$ is also independent of the state $\boldsymbol{S}$. Consider two different states $s_2$ and $\tilde{s}_2$ with $|s_2| \geq |\tilde{s}_2|$. According to (38), we have:

$$\begin{cases} H(s_2 X + Z_2|U) = \log \pi e(|s_2|^2 \alpha(s_2)P + 1) \\ H(\tilde{s}_2 X + Z_2|U) = \log \pi e(|\tilde{s}_2|^2 \alpha(\tilde{s}_2)P + 1) \end{cases} \tag{39}$$

Let $\tilde{Z}_2$ be a virtual Gaussian noise independent of $Z_2$ with zero mean and unit-variance. Therefore, we have:

$$H(\tilde{s}_2 X + Z_2|U) = H\left(\frac{\tilde{s}_2}{s_2}(s_2 X + Z_2) + \sqrt{1 - \left|\frac{\tilde{s}_2}{s_2}\right|^2} \tilde{Z}_2 \middle| U\right)$$

$$\overset{(a)}{\geq} \log\left(2^{H\left(\frac{\tilde{s}_2}{s_2}(s_2 X + Z_2)|U\right)} + 2^{H\left(\sqrt{1 - \left|\frac{\tilde{s}_2}{s_2}\right|^2} \tilde{Z}_2 \middle| U\right)}\right)$$

$$\overset{(b)}{=} \log\left(\pi e \left|\frac{\tilde{s}_2}{s_2}\right|^2 (s_2^2 \alpha(s_2)P + 1) + \pi e \left(1 - \left|\frac{\tilde{s}_2}{s_2}\right|^2\right)\right) = \log(\pi e(\tilde{s}_2^2 \alpha(s_2)P + 1))$$

$$\tag{40}$$

where the inequality (a) is due to the EPI and the equality (b) is derived based on the first equality of (39). Therefore, the function $\alpha(.)$ is decreasing. ∎

Using the outer bound of Theorem 4, one may derive more capacity results for the channel. We conclude this subsection by providing an example in this regard.

---
[3] A deterministic function $f(.): \mathbb{C} \to [0,1]$ is said to be decreasing if $\forall a, b: |a| \geq |b| \Rightarrow f(a) \leq f(b)$.

***Theorem 5)*** *Consider the two-user Gaussian fading BC (1) with degraded message sets in which the transmitter sends a common message for both users and a private message for the first user but there is no private message for the second user. Let the side information available at the transmitter be $E \equiv (E^*, E^\times)$ where $E^* \equiv (S_1, \mathbb{1}(|S_2| < |S_1|))$ and $E^\times$ is an arbitrary deterministic function of the state **S**. If the state process is i.i.d., then the capacity region is given below:*

$$\bigcup_{\beta(.),\varphi(.)} \begin{cases} (R_0, R_1) \in \mathbb{R}_+^2: \\ R_0 \leq \mathbb{E}\left[\psi\left(\frac{|S_2|^2(1-\beta(E))\varphi(E)\mathbb{1}(|S_2|<|S_1|)}{|S_2|^2\beta(E)\varphi(E)+1}\right)\right] + \mathbb{E}[\psi(|S_2|^2\varphi(E)\mathbb{1}(|S_2| \geq |S_1|))] \\ R_0 + R_1 \leq \mathbb{E}[\psi(|S_1|^2\varphi(E))] \\ R_0 + R_1 \leq \mathbb{E}[\psi(|S_1|^2\beta(E)\varphi(E)\mathbb{1}(|S_2|<|S_1|))] + \mathbb{E}\left[\psi\left(\frac{|S_2|^2(1-\beta(E))\varphi(E)\mathbb{1}(|S_2|<|S_1|)}{|S_2|^2\beta(E)\varphi(E)+1}\right)\right] \\ \qquad\qquad +\mathbb{E}[\psi(|S_2|^2\varphi(E)\mathbb{1}(|S_2| \geq |S_1|))] \end{cases}$$

(41)

where $\beta(.): \mathcal{E} \to [0,1]$ is an arbitrary deterministic function, and also $\varphi(.): \mathcal{E} \to \mathbb{R}_+$ with $\mathbb{E}[\varphi(E)] \leq P$ is the power allocation policy for the transmitter.

*Proof of Theorem 5)* Consider the achievable rate region (3). Let $\beta^*(.): \mathcal{E} \to [0,1]$ be an arbitrary deterministic function. Define the functions $\alpha(.)$ and $\beta(.)$ as follows:

$$\alpha(.) \equiv 0, \qquad \beta(E) \equiv \begin{cases} \beta^*(E) & \text{if } \mathbb{1}(|S_2|<|S_1|) = 1 \\ 0 & \text{if } \mathbb{1}(|S_2|<|S_1|) = 0 \end{cases}$$

(42)

By setting $R_2 = 0$, and also $\alpha(.)$ and $\beta(.)$ given by (42) in the rate region (3), we obtain the achievability of (41) if it is evaluated by $\beta^*(E)$. To prove the converse part, consider the outer bound in (14). By setting $R_2 = 0$, we see that $4^{th}$ constraint of this bound is redundant. Moreover, its first constraint is optimized for $\alpha(.) \equiv 0$ which yields the second constraint of the rate region (41). Since the state process is i.i.d., according to Theorem 4, one can restrict the function $\beta(.)$ in the outer bound (14) to depend only on $(S_1, E)$. For the side information in Theorem 5, we have $(S_1, E) \cong E$ because $S_1$ is a component of $E$. Accordingly, $\beta(.)$ in (14) can be restricted to depend on $E$. Thus, $2^{nd}$ and $3^{th}$ constraints of the outer bound (14) coincide with $1^{st}$ and $3^{th}$ constraints of the rate region (41), respectively. The proof is complete. ∎

### *Secrecy Capacity of the Gaussian Fading BC*

Now, we intend to follow the same approach to study the Gaussian fading BC (1) with common and confidential messages. We establish inner and outer bounds on the secrecy capacity region of the channel with partial CSIT. For the case where channel state information is perfectly known at the transmitter, our inner and outer bounds coincide which yield a full characterization of the secrecy capacity region. This new capacity result encompasses several results previously obtained for the fading BC and the fading wiretap channel using the analysis of parallel channels, specifically those given in [9-12]. For the case without common message, we also derive the secrecy capacity region when the transmitter has access only to the degradedness ordering of the channel.

Reza K. Farsani, 2013

First, we propose an achievable rate region for the channel.

**Proposition 2)** *Define the rate region* $\Re_{i \to sec}^{GFBC}$ *as follows:*

$$\Re_{i \to sec}^{GFBC} \triangleq \bigcup_{\substack{\alpha(.),\beta(.) \\ \varphi(.)}} \left\{ \begin{aligned} &(R_0, R_1, R_2) \in \mathbb{R}_+^3: \\ &R_0 \leq \min\left\{\mathbb{E}\left[\psi\left(\frac{|S_1|^2(1-\alpha(E)-\beta(E))\varphi(E)}{|S_1|^2(\alpha(E)+\beta(E))\varphi(E)+1}\right)\right], \mathbb{E}\left[\psi\left(\frac{|S_2|^2(1-\alpha(E)-\beta(E))\varphi(E)}{|S_2|^2(\alpha(E)+\beta(E))\varphi(E)+1}\right)\right]\right\} \\ &R_1 \leq \left[\mathbb{E}\left[\psi\left(\frac{|S_1|^2\beta(E)\varphi(E)}{|S_1|^2\alpha(E)\varphi(E)+1}\right)\right] - \mathbb{E}[\psi(|S_2|^2\beta(E)\varphi(E))]\right]_+ \\ &R_2 \leq \left[\mathbb{E}\left[\psi\left(\frac{|S_2|^2\alpha(E)\varphi(E)}{|S_2|^2\beta(E)\varphi(E)+1}\right)\right] - \mathbb{E}[\psi(|S_1|^2\alpha(E)\varphi(E))]\right]_+ \end{aligned} \right\}$$

(43)

*where* $\varphi(.): \mathcal{E} \to \mathbb{R}_+$ *is a power allocation policy function for the transmitter with* $\mathbb{E}[\varphi(E)] \leq P$ *and also* $\alpha(.): \mathcal{E} \to [0,1]$ *and* $\beta(.): \mathcal{E} \to [0,1]$ *are two arbitrary deterministic functions with* $\alpha(e) + \beta(e) \leq 1$ *for all* $e \in \mathcal{E}$. *The set* $\Re_{i \to sec}^{GFBC}$ *constitutes an inner bound on the secrecy capacity region of the two-user Gaussian fading BC* (1) *with common and confidential messages.*

*Proof of Proposition 2)* We utilize the achievable rate region which was derived in [21, Th. 1] for the two-user BC with common and confidential messages. This region can be adapted the Gaussian fading channel (1) as follows[4]:

$$\bigcup_{\substack{P_{WUV}, \\ f(.): X = f(W,U,V,E) \\ \mathbb{E}[|X|^2] \leq P}} \left\{ \begin{aligned} &(R_0, R_1, R_2) \in \mathbb{R}_+^3: \\ &R_0 \leq \min\{I(W; Y_1|\mathbf{S}), I(W; Y_2|\mathbf{S})\} \\ &R_1 \leq [I(U; Y_1|W, \mathbf{S}) - I(U; Y_2, V|W, \mathbf{S})]_+ \\ &R_2 \leq [I(V; Y_2|W, \mathbf{S}) - I(V; Y_1, U|W, \mathbf{S})]_+ \end{aligned} \right\}$$

(44)

Now, to derive (43) it is sufficient to evaluate the rate region (44) using the signaling given in (4). ∎

We next establish an outer bound on the secrecy capacity region of the channel. For this purpose, we make use of the outer bound given in [21, Th. 2] for the discrete BC with common and confidential messages. This outer bound can be adapted for the secrecy capacity region of the Gaussian fading channel (1) as follows[5]:

$$\bigcup_{\substack{P_{X|E}P_{WUV|XS} \\ \mathbb{E}[|X|^2] \leq P}} \left\{ \begin{aligned} &(R_0, R_1, R_2) \in \mathbb{R}_+^3: \\ &R_0 \leq \min\{I(W; Y_1|\mathbf{S}), I(W; Y_2|\mathbf{S})\} \\ &R_1 \leq [I(U; Y_1|V, W, \mathbf{S}) - I(U; Y_2|V, W, \mathbf{S})]_+ \\ &R_2 \leq [I(V; Y_2|U, W, \mathbf{S}) - I(V; Y_1|U, W, \mathbf{S})]_+ \end{aligned} \right\}$$

(45)

This result is derived in spirit similar to the adaptation of the UV-outer bound for the fading channel given in Lemma 2. The details are omitted for brevity. As we see, the outer bound (45) should be optimized over three auxiliary random variables, i.e., $U, V,$ and $W$, that seems to be a rather difficult problem. Nonetheless, in the following theorem by a subtle way, we put this optimization in connection to the evaluation of the UV-outer bound given in Theorem 1 and solve the problem.

**Theorem 6)** *Define the rate region* $\Re_{o \to sec}^{GFBC}$ *as follows:*

---

[4] In fact, the inner bound of [21, Th. 1] is given for the capacity-equivocation region of the BC and the bound (44) is deduced by specializing it for the secrecy capacity region.
[5] The outer bound of [21, Th. 2] is given for the capacity-equivocation region of the discrete BC. To extract the bound (45), beside adaptation for the fading channel, we need to first specialize [21, Th. 2] for the secrecy capacity region. Two additional constraints on $R_1$ and $R_2$ can also be extracted from [21, Th. 2]; however, those given in (45) are sufficient for our purposes.

$$\mathfrak{R}_{o\to sec}^{GFBC} \triangleq \bigcup_{\substack{\alpha(.),\beta(.) \\ \varphi(.)}} \left\{ \begin{array}{l} (R_0, R_1, R_2) \in \mathbb{R}_+^3: \\ R_0 \leq \min \left\{ \begin{array}{l} \mathbb{E}\left[\psi\left(\frac{|S_1|^2(1-\alpha(\boldsymbol{S}))\varphi(E)\mathbb{1}(|S_1|<|S_2|)}{|S_1|^2\alpha(\boldsymbol{S})\varphi(E)+1}\right)\right] + \mathbb{E}\left[\psi\left(\frac{|S_1|^2(1-\beta(\boldsymbol{S}))\varphi(E)\mathbb{1}(|S_2|\leq|S_1|)}{|S_1|^2\beta(\boldsymbol{S})\varphi(E)+1}\right)\right] \\ \mathbb{E}\left[\psi\left(\frac{|S_2|^2(1-\alpha(\boldsymbol{S}))\varphi(E)\mathbb{1}(|S_1|<|S_2|)}{|S_2|^2\alpha(\boldsymbol{S})\varphi(E)+1}\right)\right] + \mathbb{E}\left[\psi\left(\frac{|S_2|^2(1-\beta(\boldsymbol{S}))\varphi(E)\mathbb{1}(|S_2|\leq|S_1|)}{|S_2|^2\beta(\boldsymbol{S})\varphi(E)+1}\right)\right] \end{array} \right\}, \\ R_1 \leq \mathbb{E}\left[\left(\psi(|S_1|^2\beta(\boldsymbol{S})\varphi(E)) - \psi(|S_2|^2\beta(\boldsymbol{S})\varphi(E))\right)\mathbb{1}(|S_2|\leq|S_1|)\right] \\ R_2 \leq \mathbb{E}\left[\left(\psi(|S_2|^2\alpha(\boldsymbol{S})\varphi(E)) - \psi(|S_1|^2\alpha(\boldsymbol{S})\varphi(E))\right)\mathbb{1}(|S_1|<|S_2|)\right] \end{array} \right\}$$

(46)

where $\varphi(.): \mathcal{E} \to \mathbb{R}_+$ *is a power allocation policy function for the transmitter with* $\mathbb{E}[\varphi(E)] \leq P$ *and also* $\alpha(.): \mathbb{C}^2 \to [0,1]$ *and* $\beta(.): \mathbb{C}^2 \to [0,1]$ *are two arbitrary deterministic functions. The set* $\mathfrak{R}_{o\to sec}^{GFBC}$ *constitutes an outer bound on the secrecy capacity region of the two-user Gaussian fading BC (1) with common and confidential messages.*

*Proof of Theorem 6)* Consider the outer bound in (45). We prove that it does not include any point outside of the rate region (46). Define new auxiliary random variables $\bar{U}$ and $\bar{V}$ as follows:

$$\bar{U} \triangleq (U, W), \qquad \bar{V} \triangleq (V, W)$$

(47)

Now, we can write:

$$R_0 \leq I(W; Y_1|\boldsymbol{S}) = \int_{|S_1|\leq|S_2|} P_{\boldsymbol{S}}(\boldsymbol{s})I(W; Y_1|\boldsymbol{s}) + \int_{|S_2|\leq|S_1|} P_{\boldsymbol{S}}(\boldsymbol{s})I(W; Y_1|\boldsymbol{s})$$
$$\leq \int_{|S_1|\leq|S_2|} P_{\boldsymbol{S}}(\boldsymbol{s})I(U, W; Y_1|\boldsymbol{s}) + \int_{|S_2|\leq|S_1|} P_{\boldsymbol{S}}(\boldsymbol{s})I(V, W; Y_1|\boldsymbol{s})$$
$$= \int_{|S_1|\leq|S_2|} P_{\boldsymbol{S}}(\boldsymbol{s})I(\bar{U}; Y_1|\boldsymbol{s}) + \int_{|S_2|\leq|S_1|} P_{\boldsymbol{S}}(\boldsymbol{s})I(\bar{V}; Y_1|\boldsymbol{s})$$

(48)

Similarly, we have:

$$R_0 \leq I(W; Y_2|\boldsymbol{S}) \leq \int_{|S_1|\leq|S_2|} P_{\boldsymbol{S}}(\boldsymbol{s})I(\bar{U}; Y_2|\boldsymbol{s}) + \int_{|S_2|\leq|S_1|} P_{\boldsymbol{S}}(\boldsymbol{s})I(\bar{V}; Y_2|\boldsymbol{s})$$

(49)

Let us carefully examine the derivations (48) and (49). We have divided the state space into two events $\{|S_1| \leq |S_2|\}$ and $\{|S_2| \leq |S_1|\}$. Then, for the case of $\{|S_1| \leq |S_2|\}$ the auxiliary $W$ is enhanced to $\bar{U}$ by adding $U$, and for the case of $\{|S_2| \leq |S_1|\}$ it is enhanced to $\bar{V}$ by adding $V$. As we see later, this is an optimal assignment for several special cases.

Also, for the rate $R_1$ one can write:

$$R_1 \leq [I(U; Y_1|V, W, \boldsymbol{S}) - I(U; Y_2|V, W, \boldsymbol{S})]_+$$
$$= \left[\int_{|S_2|\leq|S_1|} P_{\boldsymbol{S}}(\boldsymbol{s})\big(I(U; Y_1|V, W, \boldsymbol{s}) - I(U; Y_2|V, W, \boldsymbol{s})\big) + \int_{|S_1|\leq|S_2|} P_{\boldsymbol{S}}(\boldsymbol{s})\big(I(U; Y_1|V, W, \boldsymbol{s}) - I(U; Y_2|V, W, \boldsymbol{s})\big)\right]_+$$
$$\overset{(a)}{\leq} \int_{|S_2|\leq|S_1|} P_{\boldsymbol{S}}(\boldsymbol{s})\big(I(U; Y_1|V, W, \boldsymbol{s}) - I(U; Y_2|V, W, \boldsymbol{s})\big)$$
$$= \int_{|S_2|\leq|S_1|} P_{\boldsymbol{S}}(\boldsymbol{s})\left(\big(I(X, U; Y_1|V, W, \boldsymbol{s}) - I(X, U; Y_2|V, W, \boldsymbol{s})\big) + \big(I(X; Y_2|U, V, W, \boldsymbol{s}) - I(X; Y_1|U, V, W, \boldsymbol{s})\big)\right)$$

$$\overset{(b)}{\leq} \int_{|S_2|\leq|S_1|} P_S(s)\big(I(X,U;Y_1|V,W,s) - I(X,U;Y_2|V,W,s)\big)$$

$$\overset{(c)}{=} \int_{|S_2|\leq|S_1|} P_S(s)\big(I(X;Y_1|\bar{V},s) - I(X;Y_2|\bar{V},s)\big)$$

$$= \int_{|S_2|\leq|S_1|} P_S(s)\big(H(Y_1|\bar{V},s) - H(Y_2|\bar{V},s) - H(Z_1) + H(Z_2)\big)$$

$$= \int_{|S_2|\leq|S_1|} P_S(s)\big(H(Y_1|\bar{V},s) - H(Y_2|\bar{V},s)\big)$$

(50)

where the inequality (a) holds because when $|S_1| \leq |S_2|$, the output $Y_1$ is a degraded version of $Y_2$ and thereby the second integral in the left side of (a) is negative; the inequality (b) holds because when $|S_2| \leq |S_1|$, the output $Y_2$ is a degraded version of $Y_1$ and thereby $I(X;Y_2|U,V,W,s) \leq I(X;Y_1|U,V,W,s)$; and lastly, the equality (c) holds because $U,V,W \to X \to Y_1,Y_2$ forms a Markov chain.

Symmetrically, we can obtain:

$$R_2 \leq [I(V;Y_2|U,W,S) - I(V;Y_1|U,W,S)]_+ \leq \int_{|S_1|\leq|S_2|} P_S(s)\big(H(Y_2|\bar{U},s) - H(Y_1|\bar{U},s)\big)$$

(51)

Then, consider the right side of (48)-(51). Similar to the derivations (24)-(26), one can deduce that for $s \in \{|S_1| \leq |S_2|\}$ there exist $0 \leq \alpha(s) \leq 1$ so that:

$$\begin{cases} H(Y_2|\bar{U},s) = \log \pi e(|s_2|^2\alpha(s)\varphi(e) + 1) \\ H(Y_1|\bar{U},s) \geq \log \pi e(|s_1|^2\alpha(s)\varphi(e) + 1) \end{cases}$$

(52)

Therefore, we have:

$$\int_{|S_1|\leq|S_2|} P_S(s)I(\bar{U};Y_1|s) \leq \mathbb{E}\left[\psi\left(\frac{|S_1|^2(1-\alpha(S))\varphi(E)\mathbb{1}(|S_1|<|S_2|)}{|S_1|^2\alpha(S)\varphi(E) + 1}\right)\right]$$

$$\int_{|S_1|\leq|S_2|} P_S(s)I(\bar{U};Y_2|s) \leq \mathbb{E}\left[\psi\left(\frac{|S_2|^2(1-\alpha(S))\varphi(E)\mathbb{1}(|S_1|<|S_2|)}{|S_2|^2\alpha(S)\varphi(E) + 1}\right)\right]$$

$$\int_{|S_1|\leq|S_2|} P_S(s)\big(H(Y_2|\bar{U},s) - H(Y_1|\bar{U},s)\big) \leq \mathbb{E}\left[\big(\psi(|S_2|^2\alpha(S)\varphi(E)) - \psi(|S_1|^2\alpha(S)\varphi(E))\big)\mathbb{1}(|S_1|<|S_2|)\right]$$

(53)

Symmetrically, we can deduce that there exist $0 \leq \beta(S) \leq 1$ so that:

$$\int_{|S_2|\leq|S_1|} P_S(s)I(\bar{V};Y_1|s) \leq \mathbb{E}\left[\psi\left(\frac{|S_1|^2(1-\beta(S))\varphi(E)\mathbb{1}(|S_2|\leq|S_1|)}{|S_1|^2\beta(S)\varphi(E) + 1}\right)\right]$$

$$\int_{|S_2|\leq|S_1|} P_S(s)I(\bar{V};Y_2|s) \leq \mathbb{E}\left[\psi\left(\frac{|S_2|^2(1-\beta(S))\varphi(E)\mathbb{1}(|S_2|\leq|S_1|)}{|S_2|^2\beta(S)\varphi(E) + 1}\right)\right]$$

$$\int_{|S_2|\leq|S_1|} P_S(s)\big(H(Y_1|\bar{V},s) - H(Y_2|\bar{V},s)\big) \leq \mathbb{E}\left[\big(\psi(|S_1|^2\beta(S)\varphi(E)) - \psi(|S_2|^2\beta(S)\varphi(E))\big)\mathbb{1}(|S_2|\leq|S_1|)\right]$$

(54)

By substituting (53) and (54) in (48)-(51), we derive the outer bound (46). The proof is thus complete. ∎

***Corollary 1)*** *Consider the two-user Gaussian fading BC without common message. The following constitutes an outer bound on the secrecy capacity region:*

$$\bigcup_{\varphi(.)} \begin{Bmatrix} (R_1, R_2) \in \mathbb{R}_+^2 : \\ R_1 \leq \mathbb{E}\left[\left(\psi(|S_1|^2 \varphi(E)) - \psi(|S_2|^2 \varphi(E))\right) \mathbb{1}(|S_2| \leq |S_1|)\right] \\ R_2 \leq \mathbb{E}\left[\left(\psi(|S_2|^2 \varphi(E)) - \psi(|S_1|^2 \varphi(E))\right) \mathbb{1}(|S_1| < |S_2|)\right] \end{Bmatrix}$$

(55)

where $\varphi(.): \mathcal{E} \to \mathbb{R}_+$ is a power allocation policy function for the transmitter with $\mathbb{E}[\varphi(E)] \leq P$.

*Proof of Corollary 1)* This is directly derived from the outer bound $\mathfrak{R}_{o \to sec}^{GFBC}$ in (46) if we consider only the constraints given on the rates $R_1$ and $R_2$. In fact, without considering the constraint on $R_0$, the bound $\mathfrak{R}_{o \to sec}^{GFBC}$ is optimized for $\alpha(.) \equiv \beta(.) \equiv 1$. ∎

We then prove that for the case of perfect CSIT, the inner bound (43) and the outer bound (46) coincide which yields the secrecy capacity region explicitly. This result is given in the next theorem.

**Theorem 7)** *Consider the two-user fading BC (1) with common and confidential messages. Assume that the state information is perfectly available at the transmitter, i.e., $E \equiv S$. The secrecy capacity region is given by:*

$$\bigcup_{\substack{\alpha(.), \beta(.) \\ \varphi(.)}} \begin{Bmatrix} (R_0, R_1, R_2) \in \mathbb{R}_+^3 : \\ R_0 \leq \min \begin{cases} \mathbb{E}\left[\psi\left(\frac{|S_1|^2(1-\alpha(S))\varphi(S)\mathbb{1}(|S_1|<|S_2|)}{|S_1|^2\alpha(S)\varphi(S)+1}\right)\right] + \mathbb{E}\left[\psi\left(\frac{|S_1|^2(1-\beta(S))\varphi(S)\mathbb{1}(|S_2|\leq|S_1|)}{|S_1|^2\beta(S)\varphi(S)+1}\right)\right], \\ \mathbb{E}\left[\psi\left(\frac{|S_2|^2(1-\alpha(S))\varphi(S)\mathbb{1}(|S_1|<|S_2|)}{|S_2|^2\alpha(S)\varphi(S)+1}\right)\right] + \mathbb{E}\left[\psi\left(\frac{|S_2|^2(1-\beta(S))\varphi(S)\mathbb{1}(|S_2|\leq|S_1|)}{|S_2|^2\beta(S)\varphi(S)+1}\right)\right] \end{cases} \\ R_1 \leq \mathbb{E}\left[\left(\psi(|S_1|^2\beta(S)\varphi(S)) - \psi(|S_2|^2\beta(S)\varphi(S))\right)\mathbb{1}(|S_2|\leq|S_1|)\right] \\ R_2 \leq \mathbb{E}\left[\left(\psi(|S_2|^2\alpha(S)\varphi(S)) - \psi(|S_1|^2\alpha(S)\varphi(S))\right)\mathbb{1}(|S_1|<|S_2|)\right] \end{Bmatrix}$$

(56)

where $\varphi(.): \mathcal{E} \to \mathbb{R}_+$ is a power allocation policy function for the transmitter with $\mathbb{E}[\varphi(E)] \leq P$ and also $\alpha(.): \mathbb{C}^2 \to [0,1]$ and $\beta(.): \mathbb{C}^2 \to [0,1]$ are two arbitrary deterministic functions.

*Proof of Theorem 7)* The proof is similar to Theorem 2. Let $\alpha^*(.): \mathbb{C}^2 \to [0,1]$ and $\beta^*(.): \mathbb{C}^2 \to [0,1]$ be two arbitrary deterministic functions. Define the deterministic functions $\alpha(.): \mathbb{C}^2 \to [0,1]$ and $\beta(.): \mathbb{C}^2 \to [0,1]$ as follows:

$$\alpha(S) \triangleq \begin{cases} \alpha^*(S) & \text{if } |S_1| < |S_2| \\ 0 & \text{if } |S_1| \geq |S_2| \end{cases}, \quad \beta(S) \triangleq \begin{cases} 0 & \text{if } |S_1| < |S_2| \\ \beta^*(S) & \text{if } |S_1| \geq |S_2| \end{cases}$$

(57)

Thereby, we have $\alpha(s) + \beta(s) \leq 1$ for all $s \in \mathbb{C}^2$. Now by substituting $\alpha(.)$ and $\beta(.)$ in the achievable rate region $\mathfrak{R}_{i \to sec}^{GFBC}$ in (43), one can see that it is equal to the rate region $\mathfrak{R}_{o \to sec}^{GFBC}$ in (46) if it is evaluated by $\alpha^*(S)$ and $\beta^*(S)$. This completes the proof. ∎

*Remarks 5:*

1. Theorem 7 contains all the results of [9-12] as special cases. Specifically, by setting $R_2 = 0$ in (56), the resultant region is optimized for $\alpha(.) \equiv 0$ and we re-derive [10, Corollary 5]. Also, by setting $R_0 = 0$ in (56), the resultant region is optimized for $\alpha(.) \equiv \beta(.) \equiv 0$ and we re-derive the results of [9, 11, 12]. Note that the results of all the latter papers are derived by resorting to the analysis of parallel channels. We have obtained a stronger result with a more concise proof.
2. Theorem 7 also completes a gap in the paper [11]. Clearly, in [11] Ekrem and Ulukus could find the secrecy capacity region of the parallel degraded BCs [11, Corollary 4] with both common and confidential messages, however, for the Gaussian fading channel the secrecy capacity region is given only for the channel without common message. In other words, for the Gaussian fading BC with both common and confidential messages the secrecy capacity region remains unresolved in [11].

While the condition of perfect CSIT is constructive to achieve a full characterization of the secrecy capacity region in Theorem 7, it is rather restrictive from practical viewpoints because in some practical scenarios it is not possible to equip the transmitter with perfect channel state information in a timely fashion. Unfortunately, for the channel with partial CSIT the inner bound (43) and the outer bound (46) may not coincide because the functions $\alpha(.)$ and $\beta(.)$ for the inner bound depend on the side information $E$ while for the outer bound they depend on the state $\mathbf{S}$. Nonetheless, as given in Corollary 1, for the channel without common message the outer bound (46) is reduced to (55) which does not include the functions $\alpha(.)$ and $\beta(.)$. As a result, we can derive the secrecy capacity region for a more general setting from the viewpoint of the CSIT quality. Specifically, we have the following theorem.

**Theorem 8)** *Consider the two-user Gaussian fading BC (1) without common message. Assume that the transmitter has access to the degradedness ordering of the channel, i.e., $E \equiv (E^*, E^\times)$ where $E^* \equiv \mathbb{1}(|S_2| < |S_1|)$ and $E^\times$ is an arbitrary deterministic function of the state $\mathbf{S}$. The secrecy capacity region is given by:*

$$\bigcup_{\varphi(.)} \begin{cases} (R_1, R_2) \in \mathbb{R}_+^2: \\ R_1 \leq \mathbb{E}\left[\left(\psi(|S_1|^2 \varphi(E)) - \psi(|S_2|^2 \varphi(E))\right) \mathbb{1}(|S_2| \leq |S_1|)\right] \\ R_2 \leq \mathbb{E}\left[\left(\psi(|S_2|^2 \varphi(E)) - \psi(|S_1|^2 \varphi(E))\right) \mathbb{1}(|S_1| \leq |S_2|)\right] \end{cases}$$

(58)

*where $\varphi(.): \mathcal{E} \to \mathbb{R}_+$ is a power allocation policy function for the transmitter with $\mathbb{E}[\varphi(E)] \leq P$.*

*Proof of Theorem 8)* Consider the achievable rate region (43). Define the functions $\alpha(.)$ and $\beta(.)$ as follows:

$$\alpha(E) \equiv \begin{cases} 1 & \text{if } \mathbb{1}(|S_1| < |S_2|) = 1 \\ 0 & \text{if } \mathbb{1}(|S_1| < |S_2|) = 0 \end{cases}, \quad \beta(E) \equiv \begin{cases} 1 & \text{if } \mathbb{1}(|S_2| < |S_1|) = 1 \\ 0 & \text{if } \mathbb{1}(|S_2| < |S_1|) = 0 \end{cases}$$

(59)

By setting $R_0 = 0$ and substituting $\alpha(.)$ and $\beta(.)$ in (43), we obtain the achievability of (58). The converse part is directly given by Corollary 1. ∎

The bounds derived in this paper for the fading channel with partial CSIT can be explicitly computed by standard arguments in convex optimization. It is also remarked that the optimum power allocation for the fading BC (without secrecy) with perfect CSIT, where the transmitter sends only private messages to the receivers, is given in [1]. Also, the optimum power allocation is derived in [31] for the case with both private and common messages. For the channel with secrecy when perfect state information is available at the transmitter, the optimum power allocation is given in [10-12].

## CONCLUSION

In this paper, we developed a new approach for analyzing wireless ergodic fading BCs with arbitrary stationary fading statistics and any arbitrary amount of CSIT. Specifically, a novel method was presented to evaluate the well-known UV-outer bound for the Gaussian fading BCs using the entropy power inequality. Several new capacity results were established which include all previous results as special cases, as well. The approach is also applicable to analyze various fading network topologies regardless of that a given network is separable into parallel sub-channels or not (specially, wireless fading interference networks [23]). This paper presented our approach for the derivation of capacity bounds. The evaluation of the derived bounds is addressed in [32].